\documentclass[preprint,double-spaced,aps,apssymb]{revtex4-1}
\usepackage{amsmath,amssymb}
\usepackage{graphicx}
\usepackage{dcolumn}
\usepackage{bm}
\usepackage{subfigure}
\newcommand{\ber}{\begin{eqnarray}}
\newcommand{\eer}{\end{eqnarray}}
\newcommand{\bea}{\begin{equation}}
\newcommand{\eea}{\end{equation}}

\begin{document}

\title{Generalized fluctuation-dissipation relation and statistics for the equilibrium of a system with conformation dependent damping }

\author{A. Bhattacharyay}
\email{a.bhattacharyay@iiserpune.ac.in}
 \affiliation{Indian Institute of Science Education and Research, Pune, India}

\date{\today}

\begin{abstract}
Liouville's theorem, based on the Hamiltonian flow (micro-canonical ensemble) for a many particle system, indicates that the (stationary) equilibrium probability distribution is a function of the Hamiltonian. A canonical ensemble corresponds to a micro-canonical one at thermodynamic limit. On the contrary, the dynamics of a single Brownian particle (BP) being explicitly non-Hamiltonian with a force and damping term in it and at the other extreme to thermodynamic limit admits the Maxwell-distribution (MD) for its velocity and Boltmann-distribution (BD) for positions (when in a potential). This is due to the fluctuation-dissipation relation (FDR), as was first introduced by Einstein, which forces the Maxwell distribution to the Brownian particles. For a structureless BP, that, this theory works is an experimentally verified fact over a century now. Considering a structured Brownian particle we will show that the BD and MD fails to ensure equilibrium. We will derive a generalized FDR on the basis of the demand of zero current on inhomogeneous space. Our FDR and resulting generalized equilibrium distributions recover the standard ones at appropriate limits.
\end{abstract}
\maketitle
\par
A small system (mesoscopic system) like a Brownian particle (BP), which is far away from thermodynamic limit, can be studied within the framework of Langevin equation \cite{lem}. Langevin equation is Newton's equation of motion for such systems with a dissipative term and a force term in it which represent the average interactions of the system with the heat-bath. For the system to equilibrate, its energy will, on average (ideally over some span of space and time), become a constant and that requires a balance between the energy gain from the bath and that dissipated away to the bath. The FDR ensures this energy balance. As was first shown by Einstein \cite{fur} in some equivalent form, the strength of the stochastic force $\sqrt{2\Gamma k_BT}$ on a BP (of considerably bigger size than the molecules of the heat bath its immersed in) is related to its damping constant $\Gamma$. Such a particle will equilibrate with the heat-bath at temperature $T$ with a Maxwellian velocity distribution $e^{-mv^2/2k_BT}$ \cite{swa}. The reason being, the equipartition of energy \cite{reif} is inbuilt in the FDR and equipartition is a consequence of the Maxwell-distribution of velocity. In the above expressions $k_B$ is Boltzmann constant, $v$ is the velocity and $m$ is the mass of the BP. Note that, the Brownian particle being an object with no structural fluctuations and the heat-bath being homogeneous everywhere, the $\Gamma$ is a constant. 
\par
Fluctuation dissipation relation in general is an expression of the dissipation (response) of the system in terms of the correlations of the stochastic forces \cite{swa,car,lit,cha}. Einstein's FDR, which was given for a conventional Brownian particle considers a constant damping constant for the particle where the underlying reality being the BP is large compared to the size of the bath degrees of freedom. For Brownian particles of the same dimensions as that of the bath degrees of freedom, a classic example has been given by Kubo \cite{kubo} considering a time dependent dissipation for a BP of comparable size of the bath degrees of freedom. The FDR is not only useful as an equilibrium condition, its the most widely used instrument to look into non-equilirium states and has been proposed in a number of ways and forms \cite{kubo,wang,eva,lip,cem}. In the present letter, with a simple model system characterized by a space dependent damping $\Gamma(x)$, we will show that a standard form of FDR and the resulting MB distribution does not ensure the most important ingredient of equilibrium i.e. zero current (consequently zero entropy production) over the associated inhomogeneous phase space of the system. We will work out a generalization of the FDR to get a distribution that ensures required zero current in equilibrium. This will immediately mean a generalization of the Maxwell-Boltzmann distribution which we recover in the standard form over a spatial average. We emphasize on the fact that in what follows we will be dealing with a system with absolutely no external force working on it and the system is immersed in a heat-bath which is uniform everywhere i.e. makes the system see the same temperature everywhere in it when in equilibrium.
\par
The results that we present in this paper are due to coordinate dependence of the dissipation of small (mesoscopic) structured objects. We will model such a system with no external or global field (force) acting on it. Not to tamper with the spatial homogeneity of the heat-bath, we will consider a conformation or internal coordinate dependence of the damping constant in our model. This is the most important step of the modelling, because, an inhomogeneous bath is inherently a non-equilibrium situation. We also have another goal in modeling the system in this particular way. As there cannot be any current over an inhomogeneous space in equilibrium, there is no reason why there should not be a current over a homogeneous space in equilibrium. Any uniform motion in a homogeneous space does not produce or destroy anything (including entropy). So, equilibrium will not be disturbed. This point is particularly miss-understood by mixing up non-equilibrium steady states with a current in the presence of a gradient with the possibility of a current over homogeneous space under equilibrium. A recent work of present author shows a particular example of such an equilibrium current in a similar model \cite{ari}. We will show that the condition of no current in the conformation space practically means the possibility of a uniform current in the bath space and this would be a general proof of the special case mentioned in \cite{ari}.
\par
Our model is of the following form
\ber
\Gamma_1(z)\dot{x}_1=-\frac{\partial U(z)}{\partial x_1} + \sqrt{2D_1(z)}\eta_1(t)\\
\Gamma_2(z)\dot{x}_2=-\frac{\partial U(z)}{\partial x_2} + \sqrt{2D_2(z)}\eta_2(t).
\eer 
We consider two particles at the positions $x_1$ and $x_2$ interacting through some potential $U(z)$. Moreover, we consider that the damping constant of the particles $\Gamma_1(z)$ and $\Gamma_2(z)$ depend on the conformation (or internal coordinate) of the system $z=x_1-x_2$. The Gaussian white noise, that the individual particles experience, have their strengths dependent on the conformation of the system because the damping constants are a function of $z$ and that is a lesson learnt from the Einstein's FDR. As usual, for a Gaussian white noises, $<\eta_i(t)>=0$ and $<\eta_1(t_1)\eta_2(t_2)>=\delta_{12}\delta(t_1-t_2)$. We will find out an explicit form for these $D_i$s from the demand of the equilibrium of our system with the heat-bath. Note that, this kind of a stochastic noise is actually different from the generally considered multiplicative noise in the sense that, the present one does not make the bath space inhomogeneous \cite{jiu}.
\par
Before going into the analysis of our model, let us have a look at the meaning of this conformation dependent damping considered. For the sake of simplicity, consider the simplest thing that Stoke's law tells us the damping constant for a spherical body of radius $R$ to be proportional to radius $R$. Thus, $R$ being the internal coordinate defining the conformation of the sphere, there is clearly a conformation dependence of the damping constant. Two such spheres of radii $R_1$ and $R_2$ being immersed in the same heat-bath will see the same temperature (in equilibrium) but different damping constant and different stochastic force strengths. In fact, if a sphere changes its radius slowly enough from $R_1$ to $R_2$ being driven by the equilibrium fluctuations of the bath, its damping constant and the stochastic force strength will also change with the radius while temperature remains the same. The slowness condition is required to make sure that equilibrium prevails as the system moves around on its conformation space. In our model we have generalized such a system.  
\par 
Considering the centre of mass (CM) of the system $x=(x_1+x_2)/2$, we can recast the above equations in the form
\ber
\dot{z}=-\frac{\Gamma_1(z)+\Gamma_2(z)}{\Gamma_1(z)\Gamma_2(z)}\frac{\partial U(z)}{\partial z} + \xi_z(t) \\
\dot{x}=\frac{\Gamma_1(z)-\Gamma_2(z)}{2\Gamma_1(z)\Gamma_2(z)}\frac{\partial U(z)}{\partial z} + \xi_x(t),
\eer
where $\xi_z=\eta_1(t)\frac{\sqrt{2D_1(z)}}{\Gamma_1(z)}-\eta_2(t)\frac{\sqrt{2D_2(z)}}{\Gamma_2(z)}$ and $\xi_x = \eta_1(t)\frac{\sqrt{D_1(z)/2}}{\Gamma_1(z)}+\eta_2(t)\frac{\sqrt{D_2(z)/2}}{\Gamma_2(z)}$. The correlations of the $\xi$'s can be easily computed using those of $\eta$'s to give us $<\xi_z(t_1)\xi_z(t_2)>=\frac{2D_1(z)}{\Gamma_1(z)^2}\delta(t_1-t_2)+\frac{2D_2(z)}{\Gamma_2(z)^2}\delta(t_1-t_2)$ and $<\xi_x(t_1)\xi_x(t_2)>=\frac{D_1(z)}{2\Gamma_1(z)^2}\delta(t_1-t_2)+\frac{D_2(z)}{2\Gamma_2(z)^2}\delta(t_1-t_2)$, where $<\xi_z(t)>=<\xi_x(t)>=0$. There also exist cross correlations between $\xi_z$ and $\xi_z$ which we do not require in the present context because of complete decoupling of the conformation dynamics from the CM one.
\par
Let us concentrate on Eq.3, which is equivalent to the motion of a Brownian particle in the $z$-space where the damping depends on the coordinate. The average velocity of the particle in the inhomogeneous (due to the presence of U(z) and $\Gamma(z)$s) conformation space is given by
\bea
<\dot{z}> = -\left < \frac{\Gamma_1(z)+\Gamma_2(z)}{\Gamma_1(z)\Gamma_2(z)}\frac{d U(z)}{d z} \right > = -\int_a^b{dzP(z)\left(\frac{\Gamma_1(z)+\Gamma_2(z)}{\Gamma_1(z)\Gamma_2(z)}\right )\frac{d U(z)}{d z}},
\eea
where $P(z)$ is the equilibrium probability distribution of the system which is independent of time and the integration limits are at two extreme ends in the $z$-space where $P(z)$ vanishes. Note the most important point here, that, the standard equilibrium probability distribution of Boltzmann form i.e. $P(z)=Ne^{-U(z)/k_BT}$ ($N$ is a normalization constant) would surely not make the average current vanish for arbitrary $\Gamma(z)$s. The average current identically vanishes with Boltzmann-distribution for constant $\Gamma$s which is a standard result.
\par
Taking insights from the constant $\Gamma$ case, let us impose a sufficient condition for the average current to vanish i.e.
\bea
\frac{dP(z)}{dz}= C P(z)\zeta(z)F(z),
\eea
where $\zeta(z)=\frac{\Gamma_1(z)+\Gamma_2(z)}{\Gamma_1(z)\Gamma_2(z)}$ which has the dimension of inverse $\Gamma$, $C$ is a dimension fixing constant and $F(z)$ is the force $-\frac{d U(z)}{d z}$. This sufficient condition immediately results in a probability distribution function
\bea
P(z)=N\exp{(C\int{dz\zeta(z)F(z)})}.
\eea
This distribution function should take the form of Boltzmann distribution at constant $\Gamma$ corresponding to which $\zeta(z)=\zeta_{const}$, if the dimension fixing constant is of the form $C=\zeta_{const} k_BT$ and we would keep this in mind. The Fokker-Planck (FP) equation corresponding to the Langevin dynamics (Eq.3) can be readily written down as
\bea
\frac{\partial P(z,t)}{\partial t} = -\frac{\partial}{\partial z}\left (\zeta(z)F(z)P(z,t)- \left [\frac{2D_1(z)}{\Gamma_1(z)^2}+\frac{2D_2(z)}{\Gamma_2(z)^2}\right ]\frac{\partial}{\partial z}P(z,t)\right ).
\eea
\par
Now, if we are to produce a stationary probability distribution of the form that in general makes $<\dot{z}>$ vanish, the square bracket term in the diffusion part of the FP equation above has to be independent of conformations which is easily achieved if $D_i(z)\propto \Gamma_i(z)^2$. The proportionality constant is easy to find out keeping in mind the condition already mentioned, that, the distribution has to converge to a Boltzmann one for a constant $\Gamma$. The proportionality constant comes out to be $k_BT/<\Gamma(z)>$. Thus, the generalized probability distribution can be written down in the form
\bea
P(z)=N\exp{\frac{\int{dz\zeta(z)F(z)}}{<\zeta(z)>k_BT}},
\eea
where $<\zeta(z)>=\frac{<\Gamma_1(z)>+<\Gamma_2(z)>}{<\Gamma_1(z)><\Gamma_2(z)>}$. 
We note on passing that the general condition of a vanishing $<\dot{z}>$ implies that the $<\dot{x}>$ is not in general zero and this is the result showing that a current in homogeneous bath space is possible at equilibrium.
\par
We would generalize our analysis at this point, keeping in mind that we are effectively dealing with a single particle scenario in the form of Eq.3 and having known the generalized structure of the FDR on the basis of our model, let us apply it to a general single particle case which is not over-damped. To that end, we consider the general Langevin dynamics of a particle as  
\bea
m\dot{v}(x)=-m\Gamma(x)v(x) -F(x) +\Gamma(x)\sqrt{\frac{2mk_BT}{<\Gamma(x)>}}\eta(t)
\eea
where $x$ denotes the relevant internal coordinate and $m$ is the corresponding reduced mass. The above mentioned equation can be obtained from our dimer model easily by setting $\Gamma_1(z)=\Gamma_2(z)$ to get rid of the involvement of the centre of mass velocity term in this equation and we adopt this method for the sake of simplicity. In such a situation the centre of mass would undergo a conventional Brownian motion. One can immediately find out the expression for the distribution function from the Generalized FP equation which would obviously come out to be 
\bea
P(v,x)=N \exp{(-\frac{m<\Gamma(x)>v(x)^2}{2\Gamma(x)k_BT})} \exp{(\int_{-\infty}^x{dx^\prime\frac{<\Gamma(x)>F(x^\prime)}{\Gamma(x^\prime)k_BT}})},
\eea
and the analogy is clear with our actual two-body overdamped system. Important point to note is, the distribution function does not factorize into position and velocity parts and that is expected of a noise having spatial dependence. The normalization constant is given by $N^{-1}=\int{ dx\exp{(\int_{-\infty}^x dx^\prime{\frac{<\Gamma(x)>F(x^\prime)}{\Gamma(x^\prime)k_BT}})}}\int{ dv(x)\exp{(-\frac{m<\Gamma(x)>v(x)^2}{2\Gamma(x)k_BT})}}$. The global average kinetic energy is going to be 
\ber\nonumber
<E_{KE}>&=&\frac{mN}{2}\int{ dx\exp{(\int_{\infty}^x{\frac{<\Gamma(x)>F(x^\prime)}{\Gamma(x^\prime)k_BT}})}}\int{ dv(x)v(x)^2\exp{(-\frac{m<\Gamma(x)>v(x)^2}{2\Gamma(x)k_BT})}}\\&=&\frac{k_BT}{2<\Gamma(x)>}\int{dx\Gamma(x)P(x)}=\frac{k_BT}{2},
\eer
where $P(x)$ is obtained by integrating out $v$ from $P(v,x)$. In the over-damped limit, we can easily work out the above mentioned Boltzmann distribution for the position which would make $<v>=<\dot{x}>=0$ following the line of our previous analysis of the dimer. This is an obvious result beyond over-damped regime as well, because, the velocity distribution has explicit zero mean everywhere in space whereas the distribution width is a function of space.
\par 
One can also work out a local time average of the energy given to the system by the bath, considering the integral of the form
\bea
E_{in}=\frac{1}{2\tau}\int_0^{\tau}{dt<\xi(x,t)v(x,t)>},
\eea
where $\xi(x,t)$ is the stochastic force and $<>$ indicates here a noise-average. From the direct solution of the Eq.10, $v(x,t)$ can be found out and $v(x,t)/2$ is the average velocity within the infinitesimal interval $dt$. Considering the average velocity in this form does away with any memory in the velocity and only the velocity developed from zero initial value is taken into consideration. The factor of half comes from the approximation of an effectively linear dependence of the velocity over time $dt$ under the action of a constant force within this interval. This average local supply of energy to the system more explicitly is going to be 
\ber\nonumber
E_{in}&=& \frac{1}{2\tau}\int_0^\tau{dt\left (\frac{2k_BT\Gamma(x)^2}{<\Gamma(x)>}\right )e^{-\Gamma(x)t}\int_0^t{dt^\prime e^{\Gamma(x)t^\prime}<\eta(t)\eta(t^\prime)>}}\\ &=& m\Gamma(x)\overline{v(x)^2} = E_{diss},
\eer
where the right hand side is the average energy lost at $x$ with the local average of squared velocity $\overline{v(x)^2}$. This energy balance is crucial and ensures equilibrium over the mesoscopic time $\tau$ (bigger than the local correlation time of the noise). This relation implies a local conservation of kinetic energy which can be checked easily by multiplying Eq.10 by $v(x)$ and taking average over the local velocity. The first and last terms on the right hand side would cancel together and the middle term will vanish because $<v(x)>=0$, which will mean the kinetic energy is locally conserved.
\par
To conclude, let us have a look at the most important implications of this generalization of the FDR and the following distribution. The probability distribution that we obtained is the essential finer structure within the MB distribution (that holds globally) which has to make the average current vanish over inhomogeneous space in equilibrium. Of course, for the average current to vanish, the system has to see its whole conformation space and the MB distribution is recovered over that time and that is why we recognize our distribution as the internal structure of MB distribution. The importance of our present result is in revealing that the barrier overcoming processes in equilibrium, in general, is not governed by the potential energy alone as is implied by the standard Boltzmann distribution. Rather, the damping actually plays a role in this barrier overcoming. This may prove to be a very important step in understanding barrier overcoming processes in polymers, proteins (resolving Levinthal's paradox) where the conformation dependence of damping is a reality in all likelihood. 
\par
The other important result of our present analysis is the general proof of a uniform transport through the heat-bath under equilibrium conditions when the symmetry of the system is broken i.e. $\Gamma_1 \neq \Gamma_2$. This demonstrates Newton's first law in the context of equilibrium with a heat-bath that in general causes damping. Its important to note that, the present analysis is potentially opening up of a new direction in the understanding of complex mesoscopic systems and can be extended to a very large extent. An experimental verification of the finer structure of the probability distribution can possibly be probed by using polymers or proteins and we hope for such results to come up in future. Considering that the expression for the potential $U(z)$ is invertible one can use the potential as the variable to write the probability distribution for position as $P(z)=Ne^{-\Phi(U)/k_BT}$ where the $\zeta[Z(U)]/<\zeta[Z(U)]>=\partial \Phi(U)/\partial U$ and the force is as usual $F=-\partial U/\partial Z$. This expression has the same structure of the Boltzmann distribution. Since, at present we have a good idea about what possibly are the forces involved and their corresponding potentials, from a distribution that deviates from standard BD, one can guess $\Phi(U)$ from curve fitting and that would give an idea as to what is the form of the conformation dependent damping involved in these cases.
\par
Another important point to be mentioned is, Eq.10 which is presented here for a homogeneous bath by considering internal coordinate and reduced mass appears to generally hold for any coordinate and individual mass of particles. But, this amounts to considering an inhomogeneous space in the sense that the damping constant and the external field present are making the space inhomogeneous. Through, the form of the present FDR, however, the temperature offered by the bath remains the same everywhere. This is a fictitious situation where damping is determined by other factors than the bath alone and there exists an FDR relating strength of the stochastic force to the damping. So, no FDR of the present form should generally be extended into such an inhomogeneous space to account for equilibrium. 
\section{Discussion on It\^o vs Stratonovich dilemma in relation to our generalization}
We will start this discussion by referring to a recent paper by Lau and Lubensky \cite{lau} which gives a beautifully concise account of these conventions for tackling stochastic differential equations. At the onset we will like to mention that, in our generalization of the FDR this controversy over the choice of It\^o vs Stratonovich does become irrelevant because of the fact that the stochastic force actually appearing in the integrals is no longer position dependent. Note that, the position dependent stochastic force does not only give rise to the problem of It\^o vs Stratonovich, it also causes a common problem of drift current in both the conventions Ito and Stratonovich which is because of the choice of the stochastic force to ensure the canonical distribution as an equilibrium distribution of the system at large times. This trouble comes alike in It\^o convention and Stratonovish convention arising from the second moment of stochastic force. This ambiguity is also overcome through our method. Let us also mention in the onset that in what follows we will always consider that the spatial dependence of the stochastic force is on the internal coordinate. Otherwise, a spatial dependence of the stochastic force makes the bath inhomogeneous and that is associated with many other problems like sustenance of this inhomogeneity in the presence of bath related drift and the entropy production associated with those processes etc which are not generally addressed while claiming equilibrium on such systems, but, are genuine issues related to equilibrium.
\par
As has been shown in the ref.\cite{lau} (and in numerous other generalization of Stratonovich convention), a stochastic differential equation of the form 
\bea
\frac{\partial x}{\partial t} = -\Gamma(x)\frac{\partial H}{\partial t} + g(x)\eta(t),
\eea 
where $\eta(t)$ is a white noise. This form of an equation is considered not enough for the system to go to equilibrium at large times, one has to add one more drift term in a generalized Stratonovich approach. The term one adds (following ref.\cite{lau}) is $f_1=(1-\alpha)g(x)g^{\prime}(x)$ (where prime indicates derivative with respect to $x$). When $\alpha=0$ the convention is It\^o, $\alpha = 1/2$ its Stratonovich and $\alpha$ any other value gives us a generalized Stratonovich convention. Its interesting to note that, even in It\^o ($\alpha = 0$), there is a need for an additional drift term which is $g(x)g^{\prime}(x)$ and the sole origin of this term is the demand of a canonical distribution in equilibrium for the system to remain thermodynamically correct at the extreme other end of thermodynamic limit! This drift current is a result of considering the stochastic force as $g^2(x)=2\Gamma(x)k_BT$ to ensure canonical distribution and this problem is not there if $g(x)$ is a constant resulting in an effectively constant-width stochastic contribution to the position fluctuations.
\par
There are two crucial ingredients required for ensuring equilibrium at large times in the above mentioned methods, 1. the current is made zero over inhomogeneous space to avoid entropy production and maintaining microscopic detailed balance and 2. the probability distribution has to have a canonical form for equilibrium. The price paid is in requiring bath driven drift which is done away with by adding an extra drift term in the model and thereby essentially requiring the noise to be in a form consisting of a deterministic and a stochastic term. The second point is even more interesting, it shows that the relation required for equilibrium i.e. $g^2(x)=2\Gamma(x)k_BT$ is local and the system with inhomogeneity over space is not required to see the whole space over which it will actually equilibrate to get the strength of the stochastic force fixed for equilibrium. We argue that, a simple Brownian particle when confined by a potential gets a canonical equilibrium distribution because of the fact that it sees a uniform bath and thus the locality is not an issue. But, for a particle which sees a non-uniform heat bath and has to equilibrate over a finite region being bound to a potential, must see the whole space available to it to equilibrate.  
\par
In our modification of the expression of the stochastic force for the equilibrium of the system we have relaxed the above mentioned condition of canonical probability distribution keeping the other one of requiring zero current there. We have done so based on the conviction that the thermodynamics is not exactly the same away from thermodynamic limit. The most crucial advantage of this method is that the stochastic contribution to the position fluctuations (at the over-damped limit) of the system is a constant and has the strength $2k_BT/<\Gamma(x)>$ where $<\Gamma(x)>=\int{dx\Gamma(x)P(x)}$ where $P(x)$ is the probability of $x$. This is a strength of the force which depends on the global parameter $<\Gamma(x)>$ strictly requiring the system to see the whole inhomogeneous space available to fix the equilibrium strength of the stochastic force. In this way, there is no need of a deterministic part added to the stochastic force to get rid of spurious drifts and there is no controversy of It\^o vs Stratonovich and not even It\^o and Stratonovich. For the model with inertial terms present, it can also be converted to one with an effective stochastic force with constant strength by a change of variable from $v$ to $v/\Gamma(x)$. The Probability distribution is in a form which resembles Boltzmann form more closely because of not having any space dependent amplitude to the exponential part of the probability distribution. This constant amplitude form of the probability distribution is required for the symmetry $\int{dx P(x)fLg}=\int{dx P(x)gLf}$ to exist, where $f$ and $g$ are any doubly differentiable functions and $L=-\Gamma(x)\frac{\partial H}{\partial t}\frac{\partial}{\partial x}+g(x)\frac{\partial^2}{\partial x^2}$. This symmetry ensures detailed balance of the system and to maintain this the spurious drift terms are actually added in It\^o and Stratonovich approach. The probability distribution of ours will converge to canonical distribution when internal degrees of freedom, over which the inhomogeneity exists, are integrated out. Important to note that the fixing of the constant in the strength of the equilibrium stochastic force to $<\Gamma(x)>$ is required by the recovery of equipartition of energy over homogeneous space as is shown in Eq.12. Now, which method represents what actually happens in nature has to be determined by experiments given the apparent scarcity of information for writing down a dynamics for the evolution of the stochastic force to its equilibrium value.

\end{document}